\documentstyle[aps,psfig,multicol]{revtex}
\newcommand{\bcols}{\ifpreprintsty\else\begin{multicols}{2}\fi}
\newcommand{\ecols}{\ifpreprintsty\else\end{multicols}\fi}
\begin{document}
\draft
\title{High-frequency dynamics of wave localisation}
\author{C. W. J. Beenakker and K. J. H. van Bemmel}
\address{Instituut-Lorentz, Universiteit Leiden,
P.O. Box 9506, 2300 RA Leiden, The Netherlands}
\author{P. W. Brouwer}
\address{Laboratory of Atomic and Solid State Physics,
Cornell University, Ithaca NY 14853, USA}
\date{3 September 1999}
\maketitle
\begin{abstract}
We study the effect of localisation on the propagation of a pulse through a
multi-mode disordered wave\-guide. The correlator $\langle
u(\omega_{1})u^{\ast}(\omega_{2})\rangle$ of the transmitted wave amplitude $u$
at two frequencies differing by $\delta\omega$ has for large $\delta\omega$ the
stretched exponential tail $\propto\exp(-\sqrt{\tau_{D}\delta\omega/2})$. The
time constant $\tau_{D}=L^{2}/D$ is given by the diffusion coefficient $D$,
even if the length $L$ of the wave\-guide is much greater than the localisation
length $\xi$. Localisation has the effect of multiplying the correlator by a
frequency-independent factor $\exp(-L/2\xi)$, which disappears upon breaking
time-reversal symmetry.
\end{abstract}
\bcols

The frequency spectrum of waves propagating through a random medium contains
dynamical information of interest in optics \cite{Tig99}, acoustics
\cite{Wea94}, and seismology \cite{Whi90}. A fundamental issue is how the
phenomenon of wave localisation \cite{She90} affects the dynamics. The basic
quantity is the correlation of the wave amplitude at two frequencies differing
by $\delta\omega$. A recent microwave experiment by Genack {\em et al.}
\cite{Gen99} measured this correlation for a pulse transmitted through a
wave\-guide with randomly positioned scatterers. The waves were not localised
in that experiment, because the length $L$ of the wave\-guide was less than the
localisation length $\xi$, so the correlator could be computed from the
perturbation theory for diffusive dynamics \cite{Ber94}. The characteristic
time scale in that regime is the time $\tau_{D}=L^{2}/D$ it takes to diffuse
(with diffusion coefficient $D$) from one end of the wave\-guide to the other.
According to diffusion theory, for large $\delta\omega$ the correlator decays
$\propto\exp(-\sqrt{\tau_{D}\delta\omega/2})$ with time constant $\tau_{D}$.

What happens to the high-frequency decay of the correlator if the wave\-guide
becomes longer than the localisation length? That is the question addressed in
this paper. Our prediction is that, although the correlator is suppressed by a
factor $\exp(-L/2\xi)$, the time scale for the decay remains the diffusion time
$\tau_{D}$ --- even if diffusion is only possible on length scales $\ll L$. The
exponential suppression factor disappears if time-reversal symmetry is broken
(by some magneto-optical effect). Our analytical results are based on the
formal equivalence between a frequency shift and an imaginary absorption rate,
and are supported by a numerical solution of the wave equation.

We consider the propagation of a pulse through a disordered wave\-guide of
length $L$. In the frequency domain the transmission coefficient
$t_{nm}(\omega)$ gives the ratio of the transmitted amplitude in mode $n$ to
the incident amplitude in mode $m$. (The modes are normalized to carry the same
flux.) We seek the correlator $C(\delta\omega)=\langle
t_{nm}(\omega+\delta\omega)t_{nm}^{\ast}(\omega)\rangle$. (The brackets
$\langle\cdots\rangle$ denote an average over the disorder.)  We assume that
the (positive) frequency increment $\delta\omega$ is sufficiently small
compared to $\omega$ that the mean free path $l$ and the number of modes $N$ in
the wave\-guide do not vary appreciably, and may be evaluated at the mean
frequency $\omega$ \cite{footnote1}. We also assume that $l\gg c/\omega$ (with
$c$ the wave velocity). The localisation length is then given by \cite{Bee97}
$\xi=(\beta N+2-\beta)l$, with $\beta=1(2)$ in the presence (absence) of
time-reversal symmetry. For $N\gg 1$ the localisation length is much greater
than the mean free path, so that the motion on length scales below $\xi$ is
diffusive (with diffusion coefficient $D$).

Our approach is to map the dynamic problem without absorption onto a static
problem with absorption \cite{Kly92}. The mapping is based on the analyticity
of the transmission amplitude $t_{nm}(\omega+iy)$, at complex frequency
$\omega+iy$ with $y>0$, and on the symmetry relation
$t_{nm}(\omega+iy)=t_{nm}^{\ast}(-\omega+iy)$. The product of transmission
amplitudes $t_{nm}(\omega+z)t_{nm}(-\omega+z)$ is therefore an analytic
function of $z$ in the upper half of the complex plane. If we take $z$ real,
equal to $\frac{1}{2}\delta\omega$, we obtain the product of transmission
amplitudes $t_{nm}(\omega+\frac{1}{2}\delta\omega)
t_{nm}^{\ast}(\omega-\frac{1}{2}\delta\omega)$ considered above (the difference
with $t_{nm}(\omega+\delta\omega) t_{nm}^{\ast}(\omega)$ being statistically
irrelevant for $\delta\omega\ll\omega$). If we take $z$ imaginary, equal to
$i/2\tau_{\rm a}$, we obtain the transmission probability
$T=|t_{nm}(\omega+i/2\tau_{\rm a})|^{2}$ at frequency $\omega$ and absorption
rate $1/\tau_{\rm a}$. We conclude that the correlator $C$ can be obtained from
the ensemble average of $T$ by analytic continuation to imaginary absorption
rate:
\begin{equation}
C(\delta\omega)=\langle T\rangle\;\;{\rm for}\;\;1/\tau_{\rm a}\rightarrow
-i\delta\omega.\label{CTrelation}
\end{equation}

Two remarks on this mapping: 1. The effect of absorption (with rate
$1/\tau^{\ast}$) on $C(\delta\omega)$ can be included by the substitution
$1/\tau_{\rm a}\rightarrow -i\delta\omega+1/\tau^{\ast}$. This is of importance
for comparison with experiments, but here we will for simplicity ignore this
effect. 2. Higher moments of the product ${\cal
C}=t_{nm}(\omega+\frac{1}{2}\delta\omega)
t_{nm}^{\ast}(\omega-\frac{1}{2}\delta\omega)$ are related to higher moments of
$T$ by $\langle{\cal C}^{p}\rangle=\langle T^{p}\rangle$ for $1/\tau_{\rm
a}\rightarrow -i\delta\omega$. This is not sufficient to determine the entire
probability distribution $P({\cal C})$, because moments of the form $\langle
{\cal C}^{p}{\cal C}^{\ast q}\rangle$ can not be obtained by analytic
continuation \cite{footnote2}.

To check the validity of this approach and to demonstrate how effective it is
we consider briefly the case $N=1$. A disordered single-mode wave\-guide is
equivalent to a geometry of parallel layers with random variations in
composition and thickness. Such a randomly stratified medium is studied in
seismology as a model for the subsurface of the Earth \cite{Whi90}. The
correlator of the reflection amplitudes $K(\delta\omega)=\langle
r(\omega+\delta\omega)r^{\ast}(\omega)\rangle$ has been computed in that
context by White {\em et al.} \cite{Whi87} (in the limit $L\rightarrow\infty$).
Their result was
\begin{equation}
K(\delta\omega)=(2l/c)\delta\omega\int_{0}^{\infty}dx\,
\exp[-x(2l/c)\delta\omega]\frac{x}{x-i}.\label{Kreflection}
\end{equation}
The distribution of the reflection probability $R=|r|^{2}$ through an absorbing
single-mode wave\-guide had been studied many years earlier as a problem in
radio-engineering \cite{Tut71}, with the result
\begin{equation}
\langle R\rangle=(l/c\tau_{\rm a})\int_{1}^{\infty}dz\,\exp[-(z-1)(l/c\tau_{\rm
a})]\frac{z-1}{z+1} .\label{Raverage}
\end{equation}
One readily verifies that Eqs.\ (\ref{Kreflection}) and (\ref{Raverage}) are
identical under the substitution of $1/\tau_{\rm a}$ by $-i\delta\omega$.

In a similar way one can obtain the correlator of the transmission amplitudes
by analytic continuation to imaginary absorption rate of the mean transmission
probability through an absorbing wave\-guide. The absorbing problem for $N=1$
was solved by Freilikher, Pustilnik, and Yurkevich \cite{Fre94}. That solution
will not be considered further here, since our interest is in the multi-mode
regime, relevant for the microwave experiments \cite{Gen99}. The transmission
probability in an absorbing wave\-guide with $N\gg 1$ is given by \cite{Bro98}
\begin{equation}
\langle T\rangle=\frac{l}{N\xi_{\rm a}\sinh(L/\xi_{\rm
a})}\exp\left(-\delta_{\beta,1}\frac{L}{2Nl}\right),\label{Tmulti}
\end{equation}
for absorption lengths $\xi_{\rm a}=\sqrt{D\tau_{\rm a}}$ in the range
$l\ll\xi_{\rm a}\ll\xi$. The length $L$ of the wave\-guide should be $\gg l$,
but the relative magnitude of $L$ and $\xi$ is arbitrary. Substitution of
$1/\tau_{\rm a}$ by $-i\delta\omega$ gives the correlator
\begin{equation}
C(\delta\omega)=\frac{l\sqrt{-i\tau_{D}\delta\omega}}
{NL\sinh\sqrt{-i\tau_{D}\delta\omega}}
\exp\left(-\delta_{\beta,1}\frac{L}{2Nl}\right), \label{Cmulti}
\end{equation}
where $\tau_{D}=L^{2}/D$ is the diffusion time. The range of validity of Eq.\
(\ref{Cmulti}) is $L/\xi\ll\sqrt{\tau_{D}\delta\omega}\ll L/l$, or equivalently
$D/\xi^{2}\ll\delta\omega\ll c/l$. In the diffusive regime, for $L\ll\xi$, the
correlator (\ref{Cmulti}) reduces to the known result \cite{Ber94} from
perturbation theory.

For ${\rm max}\,(D/L^{2},D/\xi^{2})\ll\delta\omega\ll c/l$ the decay of the
absolute value of the correlator is a stretched exponential,
\begin{equation}
|C|=\frac{2l}{NL}\sqrt{\tau_{D}\delta\omega}\exp
\left(-\sqrt{\case{1}{2}\tau_{D}\delta\omega}-\delta_{\beta,1}
\frac{L}{2Nl}\right).\label{Cstretched}
\end{equation}
In the localised regime, when $\xi$ becomes smaller than $L$, the onset of this
tail is pushed to higher frequencies, but it retains its functional form. The
weight of the tail is reduced by a factor $\exp(-L/2Nl)$ in the presence of
time-reversal symmetry. There is no reduction factor if time-reversal symmetry
is broken.

\begin{figure}
\centerline{
\psfig{figure=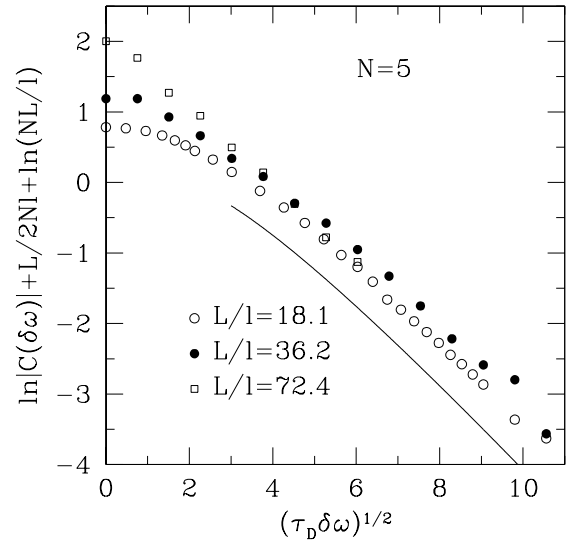,width= 8cm}}
\medskip
\caption[]{
Frequency dependence of the logarithm of the absolute value of the correlator
$C(\delta\omega)$. The data points follow from a numerical simulation for
$N=5$, the solid curve is the analytical high-frequency result
(\protect\ref{Cstretched}) for $N\gg 1$ (with $\beta=1$). The decay of the
correlator is given by the diffusive time constant $\tau_{D}=L^{2}/D$ even if
the length $L$ of the wave\-guide is greater than the localisation length
$\xi=6\,l$. The offset of about 0.6 between the numerical and analytical
results is probably a finite-$N$ effect.
\label{Cmultiplot}}
\end{figure}

To test our analytical findings we have carried out numerical simulations. The
disordered medium is modeled by a two-dimensional square lattice (lattice
constant $a$, length $L$, width $W$). The (relative) dielectric constant
$\varepsilon$ fluctuates from site to site between $1\pm\delta\varepsilon$. The
multiple scattering of a scalar wave $\Psi$ (for the case $\beta=1$) is
described by discretizing the Helmholtz equation $[\nabla^{2}+
(\omega/c)^{2}\varepsilon]\Psi=0$ and computing the transmission matrix using
the recursive Green function technique \cite{Bar91}. The mean free path $l$ is
determined from the average transmission probability $\langle{\rm
Tr}\,tt^{\dagger}\rangle=N(1+L/l)^{-1}$ in the diffusive regime \cite{Bee97}.
The correlator $C$ is obtained by averaging
$t_{nm}(\omega+\delta\omega)t_{nm}^{\ast}(\omega)$ over the mode indices $n,m$
and over different realisations of the disorder. We choose
$\omega^{2}=2\,(c/a)^{2}$, $\delta\varepsilon=0.4$, leading to $l=22.1\,a$. The
width $W=11\,a$ is kept fixed (corresponding to $N=5$), while the length $L$ is
varied in the range 400--1600~$a$. These wave\-guides are well in the localized
regime, $L/\xi$ ranging from 3--12. A large number (some $10^4$--$10^5$) of
realisations were needed to average out the statistical fluctuations, and this
restricted our simulations to a relatively small value of $N$. For the same
reason we had to limit the range of $\delta\omega$ in the data set with the
largest $L$.

Results for the absolute value of the correlator are plotted in Fig.\
\ref{Cmultiplot} (data points) and are compared with the analytical
high-frequency prediction for $N\gg 1$ (solid curve). We see from Fig.\
\ref{Cmultiplot} that the correlators for different values of $L/\xi$ converge
for large $\delta\omega$ to a curve that lies somewhat above the theoretical
prediction. The offset is about 0.6, and could be easily explained as an ${\cal
O}(1)$ uncertainty in the exponent in Eq.\ (\ref{Cmultiplot}) due to the fact
that $N$ is not $\gg 1$ in the simulation. Regardless of this offset, the
simulation confirms both analytical predictions: The stretched exponential
decay $\propto\exp(-\sqrt{\tau_{D}\delta\omega/2})$ and the exponential
suppression factor $\exp(-L/2\xi)$. We emphasize that the time constant
$\tau_{D}=L^{2}/D$ of the high-frequency decay is the diffusion time for {\em
the entire length\/} $L$ of the wave\-guide --- even though the localisation
length $\xi$ is up to a factor of 12 smaller than $L$.

We can summarize our findings by the statement that the correlator of the
transmission amplitudes {\em factorises\,} in the high-frequency regime:
$C\rightarrow f_{1}(\delta\omega)f_{2}(\xi)$. The frequency dependence of
$f_{1}$ depends on the diffusive time through the wave\-guide, even if it is
longer than the localisation length. Localisation has no effect on $f_{1}$, but
only on $f_{2}$. We can contrast this factorisation with the high-frequency
asymptotics $K\rightarrow f_{3}(\delta\omega)$ of the correlator of the
reflection amplitudes. In the corresponding absorbing problem the
high-frequency regime corresponds to an absorption length smaller than the
localisation length, so it is obvious that $K$ becomes independent of $\xi$ in
that regime. The factorisation of $C$ is less obvious. Since the localized
regime is accessible experimentally \cite{Sto99}, we believe that an
experimental test of our prediction should be feasible.

Discussions with M. B\"{u}ttiker, L. I. Glazman, K. A. Matveev, M. Pustilnik,
and P. G. Silvestrov are gratefully acknowledged. This work was supported by
the Dutch Science Foundation NWO/FOM.

\ecols

\begin{references}
\bibitem{footnote1} The length $l=\alpha_{d}l_{\rm tr}$ differs from the
transport mean free path $l_{\rm tr}$ by a dimensionality-dependent numerical
coefficient $\alpha_{d}=2,\pi/2,4/3$ for $d=1,2,3$. The diffusion coefficient
is $D=cl_{\rm tr}/d$.
\end{references}
\end{document}